\documentclass[prb,twocolumn,showpacs,preprintnumbers,amsmath,aps]{revtex4}
\usepackage{graphicx,hyperref}% Include figure files
\usepackage{bm}% bold math
\usepackage{subfigure}% bold math
\usepackage[usenames, dvipsnames]{color}

\begin{document}

\title{Avoided criticality and slow relaxation in frustrated two dimensional models}

\author{Ilya Esterlis} \email{ilyae@stanford.edu}
\affiliation{Department of Physics, Stanford University, Stanford, California 94305, USA}

\author{Steve A. Kivelson} \email{kivelson@stanford.edu}
\affiliation{Department of Physics, Stanford University, Stanford, California 94305, USA}

%\author{Srinivas Raghu} \email{sraghu@stanford.edu}
%\affiliation{Department of Physics, Stanford University, Stanford, California 94305, USA}

\author{Gilles Tarjus} \email{tarjus@lptmc.jussieu.fr}
\affiliation{LPTMC, CNRS-UMR 7600, Universit\'e Pierre et Marie Curie,
bo\^ite 121, 4 Pl. Jussieu, 75252 Paris c\'edex 05, France}

\date{\today}

\begin{abstract}
%{\bf TO BE REWRITTEN}
Frustration and the associated phenomenon of ``avoided criticality'' have been proposed as an explanation for the dramatic relaxation slowdown in glass-forming liquids. To test this, we have undertaken a Monte-Carlo study of possibly the simplest such problem, the 2-dimensional  XY model with frustration corresponding to a small flux, $f$, per plaquette.  At $f=0$, there is a Berezinskii-Kosterlitz-Thouless transition at $T^*$, but at any small but non-zero $f$, this transition is avoided, and replaced (presumably) by a vortex-ordering transition at much lower temperatures.  We thus have studied the evolution of the dynamics for small and moderate $f$ as the system is cooled from above $T^*$ to below.  While we do find strongly temperature dependent slowing of the dynamics as $T$ crosses $T^*$, and that simultaneously the dynamics becomes more complex, neither effect is anywhere nearly as dramatic as the corresponding phenomena in glass-forming liquids.  At the very least, this implies that the properties of supercooled liquids must depend on more than frustration and the existence of an avoided transition.
\end{abstract}

\pacs{????}

\maketitle

\section{Introduction}

The quest for a simple, compelling theoretical framework for understanding the spectacular dynamical phenomena exhibited ``universally'' by supercooled liquids as they approach the glass transition has been long and %contentious
arduous.\cite{kivelson-tarjus,berthier-biroli11,tarjus-review}  %Of these, the 
Surely the most dramatic of these phenomena is the super-Arrhenius temperature ($T$) dependence of the relaxation rates in the range of $T$ between the melting temperature and the glass transition temperature, $T_g$.  One set of theoretical ideas seeks to identify these dynamical phenomena with the growth of thermodynamic correlations of some sort, with the notion that geometric frustration, $f$, can be invoked to account for ultimately limiting the growth of these correlations and precluding a transition to a broken symmetry (crystalline) state.\cite{frank52,sadoc-mosseri,nelson_book}  Specifically, one concrete proposal of this variety suggests that the phenomena should be thought of as deriving from proximity to an ``avoided critical point'', $T^*$, a point at which a transition to an ``ideal'' solid phase would occur in the absence of frustration, but which is forbidden for any non-zero $f$.\cite{avoided1,avoided2,tarjus05}  

At the most optimistic level, one could then hope that any model with tunable frustration and an avoided critical point would automatically show the salient features of supercooled liquids. Besides the already mentioned spectacular increase of the relaxation time, whose temperature dependence is described by a super-Arrhenius form, the main qualitative features that one would like to reproduce are a nonexponential time dependence of the relaxation functions and the appearance of several relaxation regimes, both effects becoming more marked as one cools the liquid toward the glass transition.
To test this, we have undertaken a Monte-Carlo study of possibly the simplest such problem, the 2-dimensional  XY model with frustration corresponding to a small flux, $f$, per plaquette.  At $f=0$, there is a Berezinskii-Kosterlitz-Thouless (BKT) transition at $T^* \approx 0.89 J$, but at any small but non-zero $f$, this transition is avoided, and replaced (presumably) by a vortex-ordering transition at much lower temperatures.  We thus have studied the evolution of the (Monte-Carlo) dynamics for small $f$ as the system is cooled from above $T^*$ to below.  

While we do find strongly temperature dependent slowing of the dynamics as $T$ crosses $T^*$, and that simultaneously the dynamics becomes more complex (\textit{i.e.}, not describable as a single exponential), neither effect is anywhere nearly as dramatic as the corresponding phenomena in supercooled liquids.  At the very least, this implies that the properties of supercooled liquids must depend on more than the mere existence of an avoided transition. Conversely, it should be mentioned that in the slightly more involved example of a one-component atomic  liquid in curved space (where the curvature of space is a measure of the frustration $f$), the properties near $T^*$ much more closely resemble those of supercooled liquids, including the occurrence of a range of temperature in which super-Arrhenius slowing down is observed.\cite{sausset,sausset-nelson}

\section{Model and simulations}

The Hamiltonian of the uniformly frustrated $2$-$d$ XY model is given by\cite{teitel83}
\begin{equation}
\label{eq_hamiltonian}
H=-J\sum_{<ij>} \cos(\theta_i-\theta_j - A_{ij})
\end{equation}
where $J>0$ is the coupling constant, $\theta_i$ the angle of the XY spin at site $i$, and $<ij>$ denotes a sum over distinct pairs of nearest-neighbor sites. The bond variables $A_{ij}$ satisfy the constraint that their sum going counterclockwise around any unit cell $\mathcal C$ of the lattice is constant:
\begin{equation}
\label{eq_plaquetteflux}
\sum_{\mathcal C} A_{ij}=2\pi f\,
\end{equation}
where %by symmetry 
without loss of generality, $f$ can be restricted to the range $[0,1/2]$. %In the following 
Here we consider a square lattice. 

The 
same model has been %introduced
used  to describe an array of Josephson junctions in a uniform transverse magnetic field;\cite{teitel83,teitel11} in this case the $A_{ij}$'s can be interpreted, up to a constant, as the line integral of the vector potential along the bonds and $f$ is the number of flux quanta of magnetic field per unit cell.\cite{teitel11}

The system is \textit{frustrated} in that nonzero $f$, no matter how small, induces an irreducible density of topological defects, \textit{i.e.}, vortices %, 
all of the same sign.  %, that destroy the quasi-long-range order found in the standard 
While  the $2$-$d$ XY model (i.e. the model with $f=0$) undergoes the well known %$2$-$d$ XY model ($f=0$) below %a critical temperature 
 %$T^*$. This transition, which is in the 
BKT
 transition to a low temperature state with quasi-long-range order %occurs at a a critical temperature, 
 at $T=T^*$, %but this %universality class, is therefore \textit{avoided}. 
 for non-zero $f$ the gas of irreducible defects eliminates this transition.   These defects can crystallize,  resulting in an ordered state analogous to  the Abrikosov vortex lattice in a type-II superconducting film, but this transition  %A transition 
% to a phase in which the irreducible vortices are themselves %ordered or quasi-ordered, as 
 %crystalize, the Abrikosov vortex lattice in a type-II superconducting film, is still possible but it takes place 
 occurs at a temperature, $T_{crys}$ that is more than an order of magnitude lower than $T^*$ in the limit of small frustration $f$.\cite{fisher80,franz-teitel95,hattel95} 
The parameter $f$ %controls the irreducible density of topological defects and 
therefore quantifies the frustration that is present in the system. This frustration is  
not associated with site-dependent quenched disorder: it is {\it uniform}.
Furthermore, because the Hamiltonian in Eq. (\ref{eq_hamiltonian}) %has a gauge invariance which implies that 
is gauge invariant,  physical properties depend on the $A_{ij}$'s only through %the constant
  $f$.\cite{footnote_gauge}  
  %The frustration is therefore not due to the presence of some inhomogeneous associated with quenched disorder: it is {\it uniform}.

We have investigated the uniformly frustrated $2$-$d$ XY model described by Eq. (\ref{eq_hamiltonian}) for small to intermediate frustration $f$ both by analytical and numerical approaches.
We have performed Monte Carlo simulations for linear size $L=34$ with several values of the frustration: $f=n/L^2$ where $n$ is the irreducible number of flux quanta corresponding to $f=5/34^2$, $10/34^2$, $1/34$, $5/34$, and $13/34$. (Recall that the maximum value is $f=1/2$ %and 
which corresponds to the so-called ``fully frustrated'' XY model.\cite{teitel11})

Our Monte Carlo simulation uses single-site updates, in which each spin variable is updated according to the Metropolis method. (This corresponds to a ``model A''\cite{halperin-hohenberg} dynamics for a nonconserved order parameter.) Spins are updated by selecting a random angle with a range chosen to maintain an acceptance ratio of approximately 0.5. A single sweep corresponds to updating each spin once. In the simulations discussed here, correlation functions were measured after every sweep. 
We have made a gauge choice such that $A_{ij}=0$ for vertical bonds and $A_{ij}=2\pi f y_i$ for horizontal bonds, %with 
where $y_i$ the ordinate of site $i$. %, and $A_{ij}=0$ for vertical bonds. 
We have also studied the system in the absence of frustration ($f=0$) as a function of system size, for $L=10$ to $40$.

In the present study, we have concentrated on dynamical rather than spatial correlations.  We therefore have computed the spatially averaged spin-spin autocorrelation function %is defined as 
\begin{align}
C_{ss}%{XY}
(t) &\equiv \frac{1}{L^2}\sum_{i=1}^{L^2} <\mathbf S_i(0)\cdot \mathbf S_i(t)>  \nonumber \\
&= \frac{1}{L^2}\sum_{i=1}^{L^2} <\cos[\theta_i(0) - \theta_i(t)]> 
\label{eq_def_Cxy}
\end{align}
%The 
and the current-current autocorrelation function %s is defined as 
\begin{equation}
C_%\text{curr}
{jj} \equiv \frac{1}{L^2}\sum_{i=1}^{L^2}<\sin\Theta_i(0)\sin\Theta_i(t)>,
\label{eq_def_Ccurr}
\end{equation}
where
\begin{equation}
\Theta_i(t) \equiv \theta_i(t) - \theta_{i+\hat {\mathbf x}}(t) - A_{i,i+\hat {\mathbf x}}.
\end{equation}

\section{Avoided criticality: predictions for the dynamics}
\label{sec_predictions}

We consider the dynamical behavior of the system when it relaxes to equilibrium in the limit of small frustration $f \to 0^+$. As mentioned above, there is a large range of temperature, from around $T^*$ down to %the freezing in a lattice of frustration-induced vortices.
$T_{crys}$. In the unfrustrated model ($f=0$), the physics near and below $T^*$ can be described in terms of thermally induced defects (vortices) of both positive and negative topological charge subject to a global constraint of charge neutrality. The density of these defects decreases with decreasing temperature and below $T^*$ they only appear in dipoles formed by pairs of two nearby oppositely charged defects. The system then displays quasi-long-range order. The defect picture can be conveniently introduced through the duality transformation that modulo some approximations %(that become exact for the so-called Villain model) 
 maps the original model of XY spins into a Coulomb lattice gas with now the vortices as variables. 

For $f \neq 0$ but small, the system may locally have a tendency to behave as if unfrustrated but this cannot extend beyond an intrinsic frustration length which is of the order of $\ell = f^{-1/2}$. Indeed, frustration-induced defects must be present in addition to thermally generated ones and the irreducible density of such vortices, which all have the same topological charge of $+1$, is precisely $f$. %In two dimensions this implies that the typical distance between frustration-induced irreducible vortices is $f^{-1/2}$. 
 One may therefore expect that for some range of temperatures near and below $T^*$ and in the limit of small $f$, the frustrated model behaves as an unfrustrated one in a box of linear size of the order of $\ell= f^{-1/2}$ (with periodic boundary conditions). This is indeed what has been found for the thermodynamic properties.\cite{alba08} 
 
At still lower temperature but above the freezing transition to a vortex lattice, the system should behave as a fluid of $+1$ vortices with density $f$ in a neutralizing background, \textit{i.e.}, a ``one-component plasma''. The vortices interact with the $2$-$d$ lattice Coulomb potential $V$ which at large separation $r$ has a logarithmic dependence: %the interaction energy between two vortices is 
$%2\pi J 
V( r) %$ which goes as $
\sim -2\pi J \ln(r/a)$, with $a$ %some (irrelevant) length scale
of order a lattice constant. %, for $r$ much larger than the lattice spacing.

To make progress we first derive the finite-size scaling of the dynamics of the unfrustrated model ($f=0$) in the region below $T^*$ which is dominated by spin-wave excitations. %We consider a nonconserved Monte Carlo dynamics, which is what we have used in our computer simulations. 
The magnetization in a finite-size system of linear size $L$ %settles to an absolute value 
has a characteristic magnitude $%\vert
 M%\vert
 _{\infty} \simeq L^{-\eta(T)/2}$ with $\eta(T)=T/(2\pi J)$.\cite{footnote_M,bramwell-holdsworth93} One may then define two different time scales: a first time, $\tau_1$, to go from a magnetization of order O($1$) to a magnetization of order $%\vert
  M%\vert
  _{\infty}$ and a second one, $\tau_2$, %to relax the orientation of a magnetization of amplitude $\vert M\vert_{\infty}$. 
  which characterizes the long-time orientational fluctuations of the magnetization.

The first time $\tau_1$ is given by %the 
standard finite-size scaling %, 
to be $\tau_1\sim L^z$, with $z$ the dynamical exponent corresponding here to ``model A'', \textit{i.e.}, $z=2$. This is also obtained as the time for the spin-spin autocorrelation function to decay from $1$ to $(%\vert 
 M%\vert
 _{\infty})^2$. In this regime, the time-dependent correlation function decays as $t^{-\eta(T)/2}$ for the present dynamics, which leads indeed to
\begin{equation}
\label{eq_tau1}
\tau_1(L) \sim [L^{-\eta}]^{-\frac{2}{\eta}} \sim L^2\,,
\end{equation}
with $\tau_1$ roughly independent of the temperature.

The second time $\tau_2$ is associated with the relaxation of the magnetization angle $\theta(t)$. % in the Monte Carlo process. 
Since the process is akin to a random walk, the number of Monte Carlo steps per spin that are necessary to rotate the angle by an amount of order $2\pi$ goes as $(2\pi)^2/\langle(\delta\theta)^2\rangle$  where $\langle(\delta\theta)^2\rangle=\int d \delta \theta (\delta \theta)^2 p(\delta\theta)$ with $p(\delta \theta)$ the probability that an angular change $\delta \theta $ is made.

The probability $p(\delta\theta)$ is given by a Boltzmann factor involving the free-energy change associated with the angular change. Large angular changes are thus  strongly (exponentially) suppressed at low temperature. We therefore consider a twist of small amplitude $2\pi \delta$  in one direction across the sample, such that $\theta(x,y)=(2\pi \delta/L) x$. The associated free-energy change is given by $(J/2) \int \int dx dy [\partial \theta (x,y)]^2\sim (2\pi^2 J) \delta^2 \sim \kappa\; \delta^2$, with $\kappa$ a constant. One then estimates the mean squared angle change as
\begin{equation}
\label{eq_meansquare}
\langle(\delta\theta)^2\rangle \sim (2\pi)^2\int_0^{\delta_c} d \delta \, \delta^2 e^{-\frac{\kappa \delta^2}{T}}\, ,
\end{equation}
where $\delta_c$ is a cutoff value. At low temperature, the above expression behaves as $T^{3/2}$ and the relaxation time $\tau_2$ then scales as 
\begin{equation}
\label{eq_tau2}
\tau_2(L,T) \sim L^2\left( \frac{T^*}{T}\right)^{3/2}\,,
\end{equation}
where $L^2$ accounts for the number of spins in the system. Note that as expected $\tau_2(L,T) > \tau_1(L)$ below $T^*$.

After putting together the above results, one obtains that the spin-spin autocorrelation function %$\langle\mathbf{S}_1(0)\mathbf{.} \mathbf{S}_1(t) \rangle =\langle\cos(\theta_1(0)-\theta_1(t)) \rangle $ 
behaves in the following way:
\begin{equation}
%\begin{aligned}
\label{eq_case1}
%\langle\mathbf{S}_1(0)\mathbf{.}\mathbf{S}_1(t) \rangle
C_{ss}(t) \sim \left\{
\begin{array}{ccc}
 t^{-\frac{\eta}{2}}  &  {\rm for}  & t \ll \tau_1  \\
 \tau_1^{-\frac \eta 2} \, e^{-\frac{(t-\tau_1 )}{\tau_2 }}  &  {\rm for}  &    t \gg \tau_1.
\end{array}
\right.
%C_{ss}(t)\; &\sim\;  t^{-\frac{\eta%(T)
%}{2}}, \;\;\; {\rm for}Ê\;\; t %<
%\ll \tau_1%(L)
%\\&
%\sim\;  \tau_1^{-\frac \eta 2}%(L)
% \, e^{-\frac{(t-\tau_1%(L)
% )}{\tau_2%(L,T)
% }}, \; {\rm for}Ê\;\; t %> 
% \gg \tau_1. %(L).
%\end{aligned}
\end{equation}
Note that the exponent $\eta(T)=T/(2\pi J)$ decreases as $T$ decreases, so that the initial slope of $%\langle\mathbf{S}_1(0)\mathbf{.}\mathbf{S}_1(t) \rangle
C_{ss}(t) $ versus $\log(t)$ becomes less and less negative as $T$ decreases. In the macroscopic limit $L \rightarrow \infty$, $\tau_1(L)\rightarrow \infty$  and $%\langle\mathbf{S}_1(0)\mathbf{.}\mathbf{S}_1(t) \rangle
C_{ss}(t) \;\sim\;  t^{-\eta(T)/2}$ at all times, as required.

%On the finite size scaling of the dynamics of the $2-d$ unfrustrated XY model, see also:  S. Lepri and S. Ruffo, Europhys. Lett. \textbf{55}, 512 (2001).

What are the consequences for the frustrated model? As explained above, for a range of temperature near and below $T^*$, we expect that the behavior of the weakly frustrated ($f \ll1$) model in the thermodynamic limit is similar to that of the unfrustrated model in a finite-size box %(with periodic boundary conditions) 
of linear size of the order of the intrinsic frustration length, \textit{i.e.}, $L\approx \ell =f^{-1/2}$. The above predictions should thus apply, provided one replaces $L$ by $\ell$. For the thermodynamic quantities, Alba \textit{et al.} \cite{alba08} have found that the finite-size scaling with $L\approx f^{-1/2}$ describes their numerical simulations from $T^*\approx 0.89J$ %, which we recall is about $0.89 J$ for the square lattice, 
 down to $T=0.2J$, which is the lowest temperature they considered. %One does not know however if the range of validity is similar for the dynamics.
 However, they did not address the validity of this scaling analysis for the dynamics.

At still lower temperature (but above %the freezing in a vortex lattice
$T_{crys}$), we %have argued above that the weakly frustrated XY model should behave as a one-component plasma of density $f$.
expect to see behavior characteristic of a one-component plasma of density $f$.  In the low-temperature correlated-fluid regime of the $3$-$d$ one-component plasma, it has been observed that the dynamics can be described as an activated process with a relaxation time following an Arrhenius temperature dependence.\cite{daligault06} One %may 
might anticipate a similar behavior in 2 dimensions. 

When vortices form a dilute gas, {\it i.e.}, when the frustration $f\rightarrow 0^+$, their motion is nonetheless thermally activated at low temperature because of the periodic pinning potential associated with the (essentially) ordered spins on the underlying lattice.\cite{lobb93,franz-teitel95,hattel95}  This is analogous to the Peierls potential for dislocations in a crystal. This activation energy has been estimated by Lobb {\it et al.}\cite{lobb93} and found to be of the order of $0.19 J$ for a square lattice: this activation barrier can therefore be felt only if the system can be cooled in a disordered phase to very low temperatures significantly below $0.19 J$. (Note that $T_{crys}$ has been estimated to be around $0.045 J$.\cite{franz-teitel95,hattel95})

For a larger but still small density $f$, the one-component plasma is in a fluid phase and the activation barrier now also involves the Coulomb interaction energy between vortices. However, we expect the activation energy to be essentially independent of $f$, as in the dilute-gas regime. A crude estimate of the activation energy is given by the change in the interaction energy when displacing one vortex by a fraction of the typical separation $\ell= f^{-1/2}$. As $\ell$ is large compared to the lattice spacing, the Coulomb potential between the chosen vortex and the others can be taken as logarithmic and the change in the interaction energy is then of O($1$) irrespective of $f$. (As an illustration, consider for simplicity 3 vortices at a distance $\ell$ from each other. The cost for one vertex to pass, say, through the middle of the segment joining the two other vertices to reach another equilibrium configuration is the difference between $-2\pi J \ln(\ell/(2a))$ and $-2\pi J \ln(\ell/a)$, {\it i.e.},  $4\pi J \ln(2)$.)

For even larger frustration $f$, $\ell$ becomes of the order of the lattice spacing. The system is denser and more akin to a liquid. The activation energy should then be sensitive to the density $f$ and, as in a simple liquid,\cite{sastry00,tarjus-alba} increase with the density. This will be %further
 discussed below.

On can summarize the predicted behavior for the temperature dependence of the relaxation time $\tau(f,T)$ in the weakly frustrated XY model as follows:

1) For $T_{vl}(f) \lesssim T \lesssim T^*$, with $T%_{x}
_{vl}$ a crossover temperature whose dependence on $f$ is unknown, 
\begin{equation}
\label{eq_power_law}
\tau(f,T)\sim f^{-1}T^{-3/2}\,.
\end{equation}
2) For $T_{crys}<T \lesssim T%_{x}
_{vl}(f)$, %but still in the disordered phase, 
there is a vortex liquid regime in which
\begin{equation}
\label{eq_arrhenius}
\tau(f,T)\sim \tau_0 \, e^{\frac{\Delta_0}{T}} \, ,
\end{equation}
with $\Delta_0\sim J$ a constant activation energy that is independent of $f$ in the limit of small frustration (and possibly increases with $f$ for large enough frustration) and $\tau_0$ is a $T$-independent but possibly $f$-dependent elementary time scale. 
%\IE{Have we defined $\tau_0$?}
The first regime is controlled by spin-wave kinetics whereas the second one is due to the activated motion of the irreducible frustration-induced defects.

From the above predictions one can already see the difference with the glass-forming behavior of supercooled liquids, including that found in $2$-$d$ curved space: no generic super-Arrhenius temperature dependence is expected for the uniformly frustrated XY model; quite the contrary, a sub-Arrhenius behavior should be observed in the first temperature regime described above. %This will be discussed in more details after the presentation of the simulation results.

\section{Simulation results: the unfrustrated case}

\begin{figure}
\includegraphics[width=0.22\textwidth]{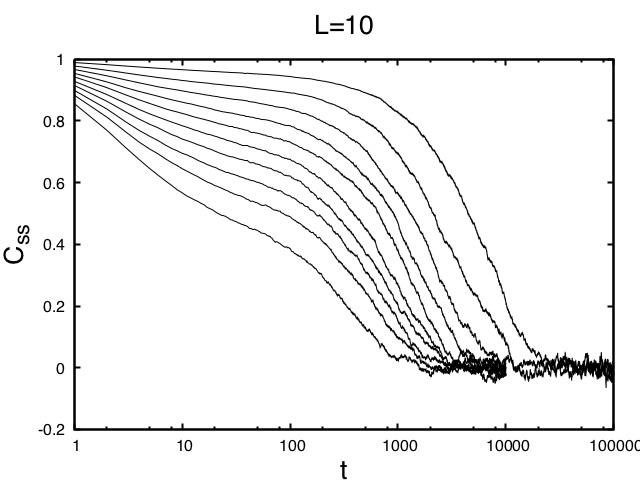}
\includegraphics[width=0.22\textwidth]{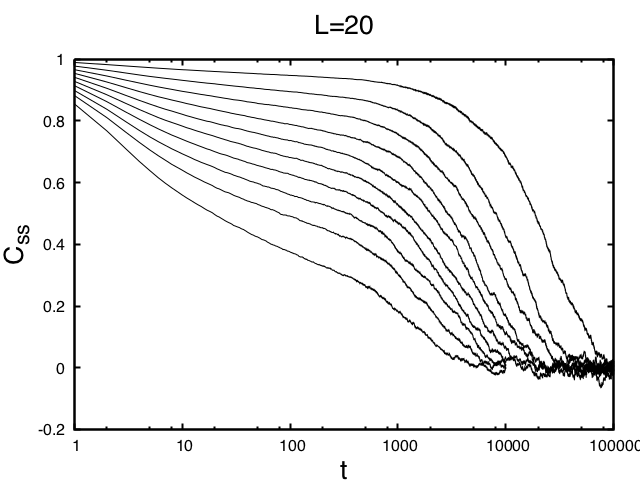}
\includegraphics[width=0.22\textwidth]{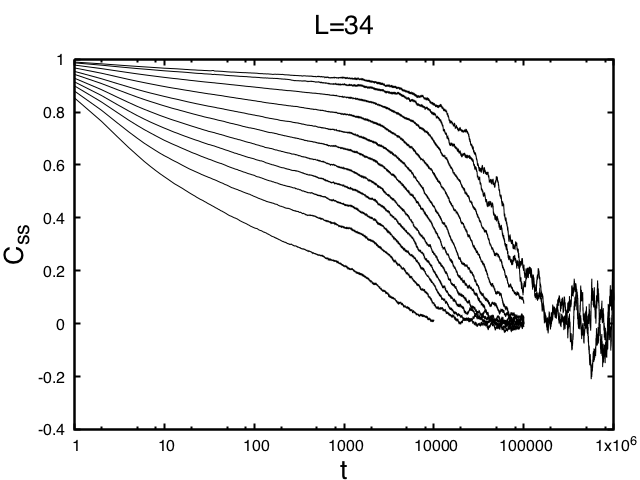}
\includegraphics[width=0.22\textwidth]{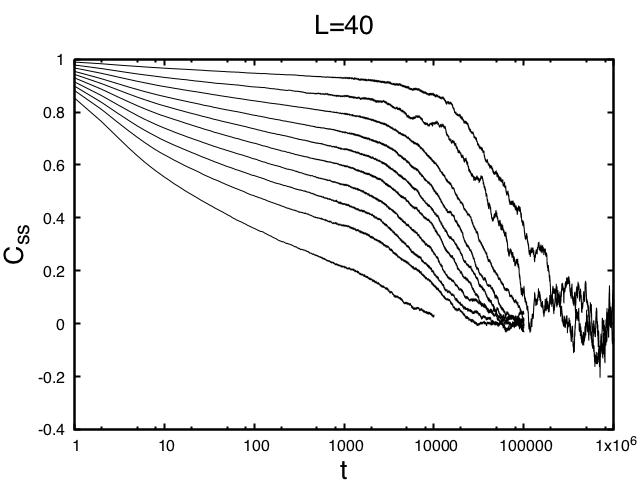}
\caption{Spin auto-correlation function $C_{ss}(t)$ versus $\log_{10}(t)$ for the unfrustrated ($f=0$) $2$-$d$ XY model at temperatures from $T/J=1$ to $0.1$ (from left to right: $T/J = $ 1.0, 0.9, 0.8, 0.7, 0.6, 0.5, 0.4, 0.3, 0.2, 0.1) for different system sizes from $L=10$ to $L=40$. System size $L=34$ includes one additional curve at $T/J=0.13$.}
\label{fig_Cxy_f0}
\end{figure}

\begin{figure}[h]
\includegraphics[width=\linewidth]{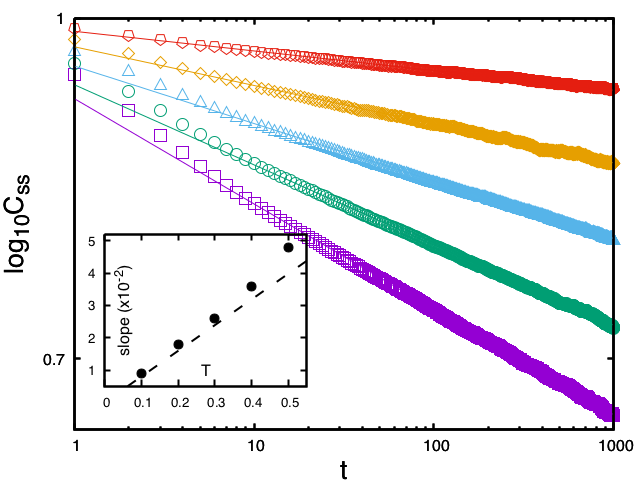}
\caption{Log-log plot of  $C_{ss}(t)$ at short times (first regime of the data in Fig.~\ref{fig_Cxy_f0})  for the unfrustrated ($f=0$) $2$-$d$ XY model with $L=40$ at several temperatures: $T/J=0.1, 0.2, 0.3, 0.4, 0.5$ (from top to bottom). Inset: Extracted slope versus temperature. The dashed line is the predicted behavior $1/(4\pi)(T/J)$.}
\label{fig_Cxy_f0_short}
\end{figure}

\begin{figure}[h]
\includegraphics[width=\linewidth]{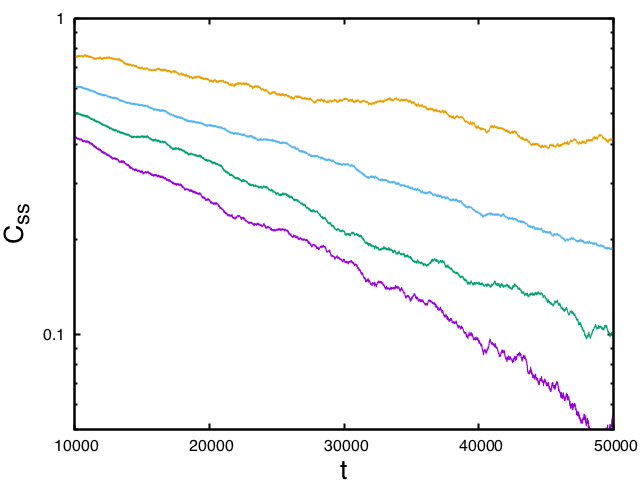}
\caption{log-linear plot of  $C_{ss}(t)$ at long times (second regime of the data in Fig.~\ref{fig_Cxy_f0})  for the unfrustrated ($f=0$) $2$-$d$ XY model with $L=40$ at several temperatures, $T/J = 0.2, 0.3, 0.4, 0.5$ from top to bottom, showing the exponential decay at long times.}
\label{fig_Cxy_f0_long}
\end{figure}

%We first consider t
The simulation results for the finite-size scaling of $C_{ss}(t)$  of the unfrustrated ($f=0$) case %. As 
are shown in Fig.~\ref{fig_Cxy_f0} for several system sizes from $L=10$ to $L=40$. % the spin auto-correlation function 
 $C_{ss}(t)$ indeed %displays two regimes. It has 
 exhibits  an exponential decay at long times, and a power-law time dependence at shorter times with an exponent that decreases as $T$ decreases. % and an exponential decay at long time. 
 In Fig.~\ref{fig_Cxy_f0_short} we show a log-log plot %of the correlation function in the first regime 
 of $C_{ss}$ for the largest system size $L=40$. %from which we extract the power-law exponent. The latter is found to go exponent
 The exponent in the power-law regime   varies linearly with temperature %with a slope that is 
 and is quantitatively compatible with the predicted value $\eta(T)/2=(T/J)/(4\pi)$, %: see the inset of Fig.~\ref{fig_Cxy_f0_short}.
 % for the largest system size under study, $L=40$.
 as shown in the inset of the figure.

\begin{figure}[h]
\includegraphics[width=\linewidth]{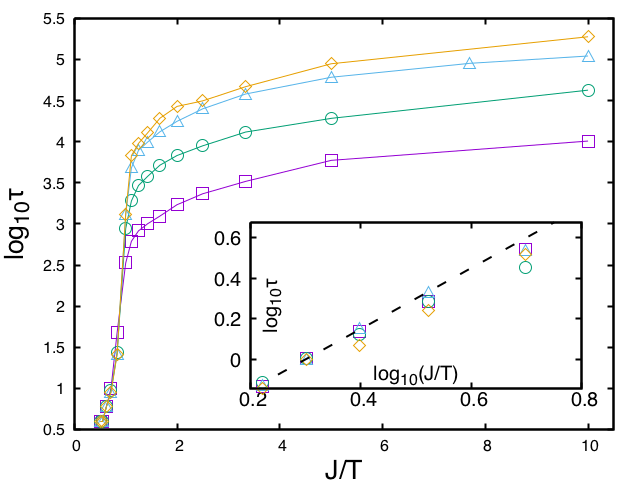}
\caption{Arrhenius plot of the $T$ dependence of the relaxation time, {\it i.e.}, $\log_{10}(\tau_2)$ versus $J/T$, for the unfrustrated ($f=0$) $2$-$d$ XY model for $L=10,20,34,40$ (from bottom to top). Inset: Plot of $\log(\tau_2)$ versus $\log(J/T)$ for $T\lesssim 0.5 J$, showing a power-law behavior compatible with $T^{-3/2}$ (displayed as the dashed straight line).}
\label{fig_tauf0}
\end{figure}

\begin{figure}[h]
\includegraphics[width=\linewidth]{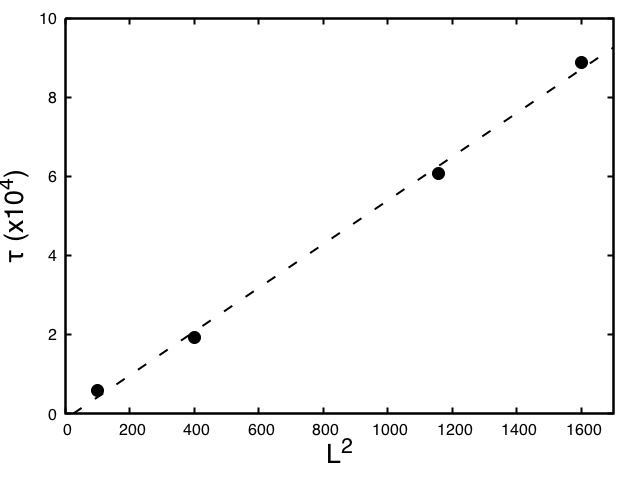}
\caption{Finite-size scaling in the unfrustrated ($f=0$) $2$-$d$ XY model for $L=10,20,34,40$:  $\tau_2$ versus $L^2$ for a low temperature $T/J=0.2$.}
\label{fig_FSS}
\end{figure}

From the exponential behavior in the long-time regime (see Fig.~\ref{fig_Cxy_f0_long}) we extract a relaxation time $\tau_2(L,T)$. This relaxation time is plotted as a function of inverse temperature  for different values of the system size $L$ in Fig.~\ref{fig_tauf0}. As shown on the log-log plot in the inset of Fig.~\ref{fig_tauf0}, $\tau_2$ %is found to 
 scales as a power-law $T^{-y}$ at low enough $T$ with the exponent $y\approx 1.5$. Furthermore, at a fixed temperature we also find that $\tau_2$ scales as $L^2$ (see Fig.~\ref{fig_FSS} for  $T=0.2 J$).

All these observations support the predictions presented in the previous section.

\section{Simulation results: the frustrated case}

We now present the results for the (equilibrium) dynamics of the frustrated model for 5 different values of frustration: $f=5/(34)^2$, $f=10/(34)^2$, $f=1/34$, $5/34$, and $13/34$. This corresponds to typical distances between irreducible vortices $\ell=f^{-1/2}$ of $\ell\approx 15.2$, $\ell\approx 10.8$, $\ell\approx 5.8$, $\ell\approx 2.6$, and $\ell\approx 1.6$, respectively (in units of the lattice spacing), and thus spans small to quite strong frustrations. The largest frustration, $f=13/34$ was  previously numerically investigated by Kim and Lee\cite{kim-lee97} as an approximation to an irrationally frustrated model with $f=$ the Golden Mean, $(3-\sqrt{5})/2$.\cite{halsey85} Slow dynamics was found and claimed to be analogous to the relaxation in supercooled liquids. This will be further discussed below.

We note that the frustrated model in the range of temperature under study has a weak dependence on system size, contrary to the unfrustrated model which is essentially always critical at and below $T^*$ (for a similar observation, see Ref.~[\onlinecite{kim-lee97}]). The results shown here are for a $34\times 34$ lattice.\cite{footnote_finitesize}

\begin{figure}
\includegraphics[width=0.22\textwidth]{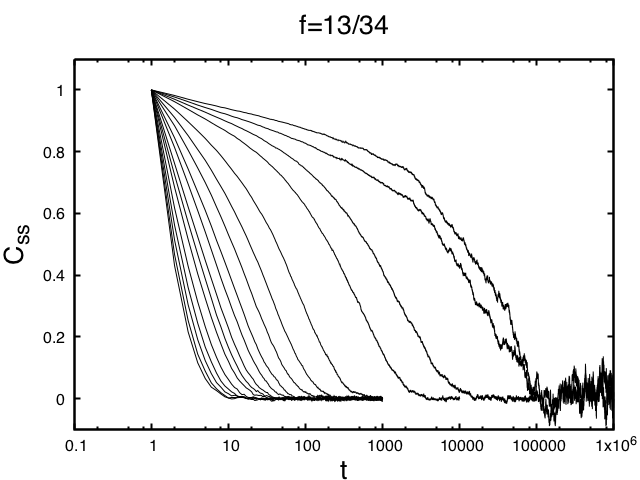}
\includegraphics[width=0.22\textwidth]{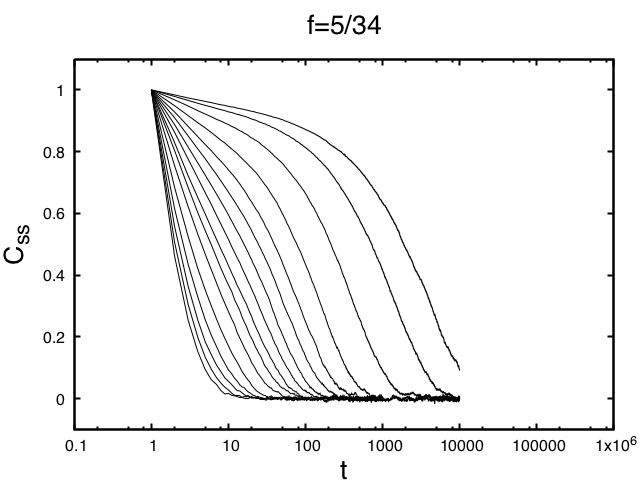}
\includegraphics[width=0.22\textwidth]{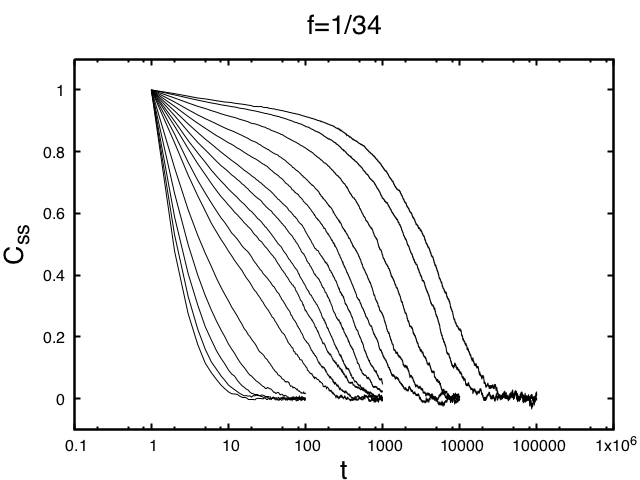}
\includegraphics[width=0.22\textwidth]{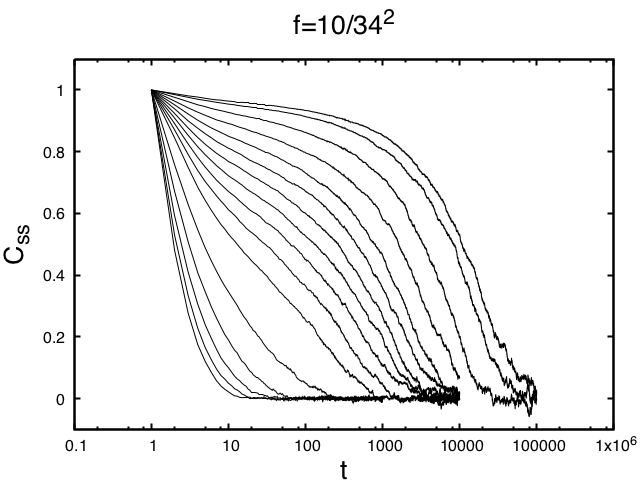}
\includegraphics[width=0.22\textwidth]{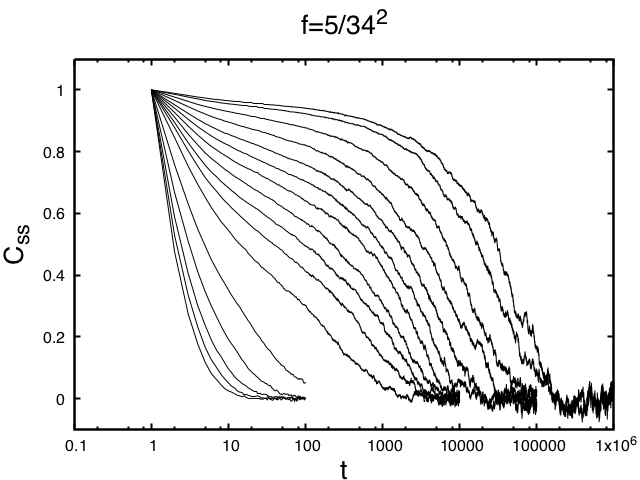}
\caption{Spin auto-correlation function  $C_{ss}(t)$ versus $\log_{10}(t)$ for the frustrated $2$-$d$ XY model at several temperatures from $T/J=2$ down to $T/J=0.1$ (from left to right: $T/J =$ 2.0, 1.8, 1.6, 1.4, 1.2, 1.0, 0.9, 0.8, 0.7, 0.6, 0.5, 0.4, 0.3, 0.2, 0.13, 0.1) for 5 frustrations: $f=13/34$, $5/34$, $1/34$, $10/(34)^2$, and $5/(34)^2$. Frustration $f=13/34$ includes one additional curve at temperature $T/J=0.16$.}
\label{fig_cxy_f}
\end{figure}

\subsection{Time-dependent correlation functions}

The spin autocorrelation function $C_{ss}(t)$ is displayed in Fig.~\ref{fig_cxy_f} as a function of the logarithm of time for the 5 frustrations and a domain of temperature from $T/J=2>T^*$ down to $T/J=0.1$. As we have %can be 
checked,  %\IE{how? should we put a reference here?}
the system stays in the disordered phase over this whole temperature regime and the BKT transition is indeed avoided. Furthermore,  as anticipated from previous studies,\cite{franz-teitel95,hattel95} there is no sign of formation of an Abrikosov %-like 
 crystal of vortices. Equilibration takes %a
  longer % time as 
  as the temperature decreases and for temperatures below $T/J=0.1$ the system falls out of equilibrium on the time scale of the simulation: the system then becomes a ``glass'', but we do not investigate this out-of-equilibrium glassy phase.

The shape of the function $C_{ss}(t)$ changes both as a function of temperature and as a function of frustration. For the three smallest frustrations,  %In the accessible temperature range, 
the behavior of $C_{ss}(t)$ appears quite similar to that of the finite-size system in the absence of frustration, as can be seen by comparing Fig.~\ref{fig_cxy_f} with Fig.~\ref{fig_Cxy_f0}. This is most visible in the short-time regime, before the appearance of a knee. In the long-time regime, the frustrated cases differ from the unfrustrated;  while the latter shows simple exponential decay at long times, the former % ones %because the 
%in that the time dependence is described by a 
exhibit stretched exponential behavior, $C_{ss}(t)\sim \exp[-(t/\tau)^\beta]$ with $\beta<1$. %, rather than by a simple one. 
However, the dependence of the stretching parameter $\beta$ on temperature is %unusual
unanticipated as $\beta$ appears to increase with decreasing temperature from $\beta\lesssim 0.5$ around $T/J=1$ to $\beta \sim 0.8$ at the lowest temperatures. This is %completely 
at odds with what  is found in supercooled liquids. 
%(***I don't understand this next sentence.***) This is confirmed by the analysis detailed below of the so-called time-temperature superposition.

For the two largest frustrations, $f=5/34$ and $f=13/34$, the behavior of $C_{ss}(t)$ changes from a simple or compressed exponential dependence at high temperature ($T>T^*$) to a two-step relaxation with the last stage being described by a stretched exponential at low temperature. This is similar to what was observed in Ref.~[\onlinecite{kim-lee97,kolahchi00}]. In these two cases, and contrary to what is observed for the smaller frustrations, the stretching parameter decreases as temperature decreases, in line with what is found in supercooled liquids: $\beta$ goes from around $1$ at $T/J=1.2$ to $0.7$ at $T/J=0.1$ for $f=5/34$ and $\beta$ goes from around $1$ at $T/J=0.9$ to $0.6$ at $T/J=0.13$ for $f=13/34$ (this is compatible with the results of Ref.~[\onlinecite{kolahchi00}]). Note however that the signature of the glassy regime in many simple glass-formers, namely the presence of a  well-developed plateau separating two relaxation regimes, is not found in the time-dependent correlation function of the present systems.

\begin{figure}[t!]
\includegraphics[width=0.22\textwidth]{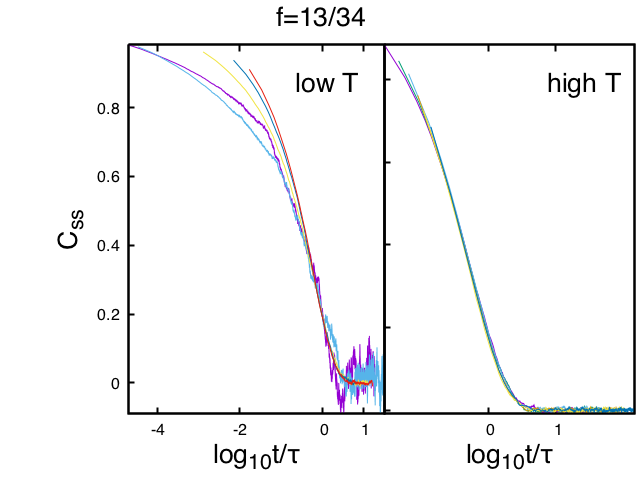}
\includegraphics[width=0.22\textwidth]{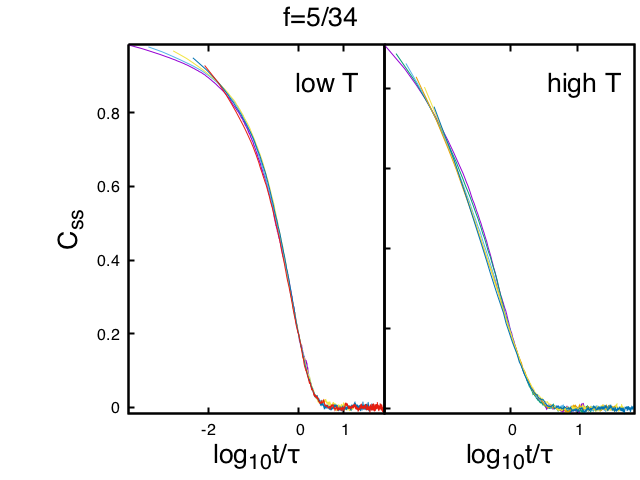}
\includegraphics[width=0.22\textwidth]{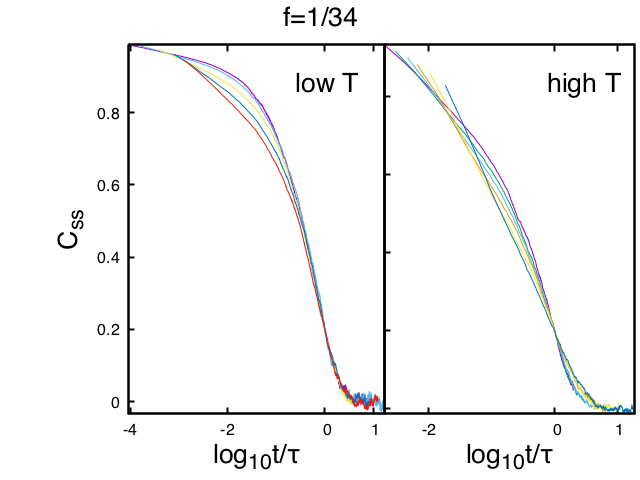}
\includegraphics[width=0.22\textwidth]{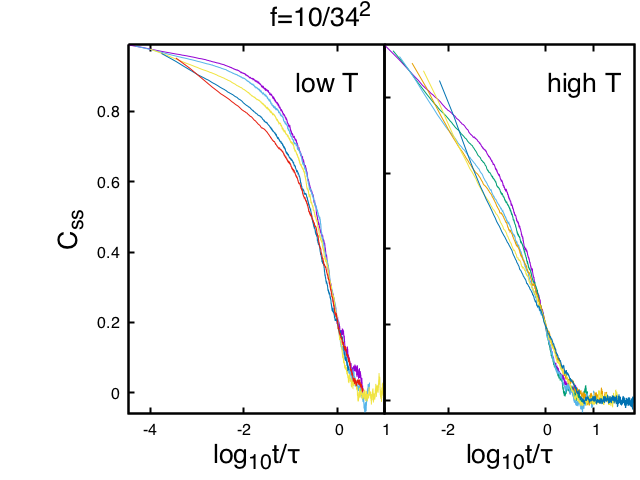}
\includegraphics[width=0.22\textwidth]{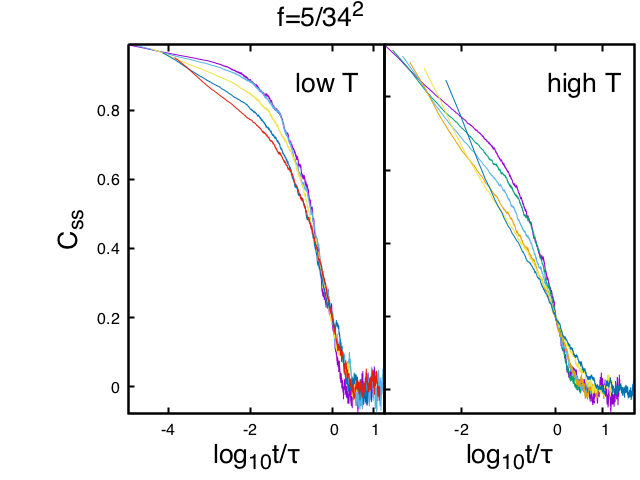}
\caption{Test of the so-called time-temperature superposition principle: same data as in Fig.~\ref{fig_cxy_f} plotted versus the logarithm of the scaled time $t/\tau(T)$. The data are divided into ``high-$T$'' and ``low-$T$'', as described in the text. While superposition is never found for the 3 smallest frustrations, it is approximately observed for the two strongest at high temperature and then violated at low temperature. }
\label{fig_cxy_f_TTS}
\end{figure}

The different types of behavior for the evolution of the shape of $C_{ss}(t)$ with temperature are also illustrated by considering plots where the time $t$ is rescaled by the relaxation time $\tau(T)$ (extracted as discussed below): see Fig.~\ref{fig_cxy_f_TTS}. The collapse on a single curve corresponds to the ``time-temperature superposition''. We have divided the range of temperature into ``high-$T$'', $T/J = 0.5-1.0$,  and ``low-$T$'', $T/J = 0.1-0.4$.  The cutoff between high and low $T$ was chosen such that high $T$ data gave the best collapse. We see that the superposition principle is never valid for the three smallest frustrations, no matter the temperature range. For the two strongest frustrations, time-temperature superposition is roughly obeyed at high-$T$ where the relaxation is essentially exponential (as discussed above this is no longer true when going to temperatures around and above $2 T^*$) but is violated at low temperature where the relaxation is stretched and the stretching parameter decreases with decreasing $T$. This latter pattern is more in line with what is usually found in glass-forming liquids.

\subsection{Temperature dependence of the relaxation time}

We have determined the relaxation time $\tau$ either as the time at which the autocorrelation function is equal to $0.2$ or as the parameter entering in the stretched-exponential description of the long-time decay. Very similar values are found through the two procedures, and below we illustrate the trend with temperature by using the second prescription (stretched-exponential fit).

\begin{figure}[h]
\includegraphics[width=\linewidth]{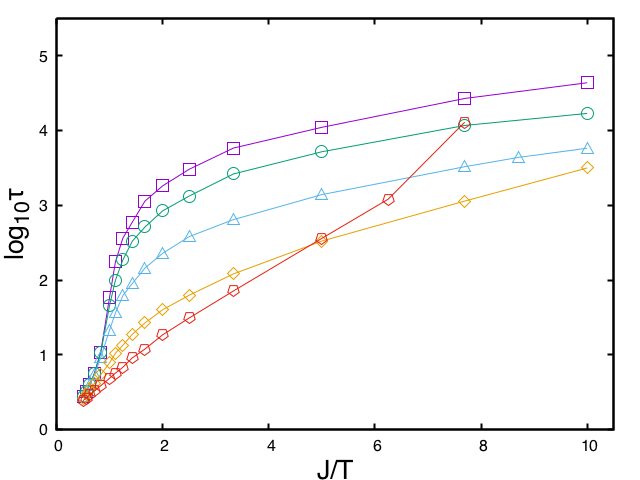}
\caption{Arrhenius plot of $\log_{10}(\tau)$ versus $J/T$ for $f=5/(34)^2, 10/(34)^2,1/34, 5/34, 13/34$ (from top to bottom).}
\label{fig_arrhenius_tau}
\end{figure}

The logarithm of the relaxation time $\tau$ is shown in Fig.~\ref{fig_arrhenius_tau} as a function of $J/T$ for the 5 frustrations. The relaxation time increases with decreasing temperature, but the dependence is quite different from what is seen for glass-forming liquids: in place of a super-Arrhenius dependence with a positive curvature on the Arrhenius plot, one finds first an opposite trend with a rapid increase followed by some form of sub-Arrhenius behavior at intermediate temperature, and finally, when accessible, a low-$T$ Arrhenius regime. 

\begin{figure}[h]
\includegraphics[width=\linewidth]{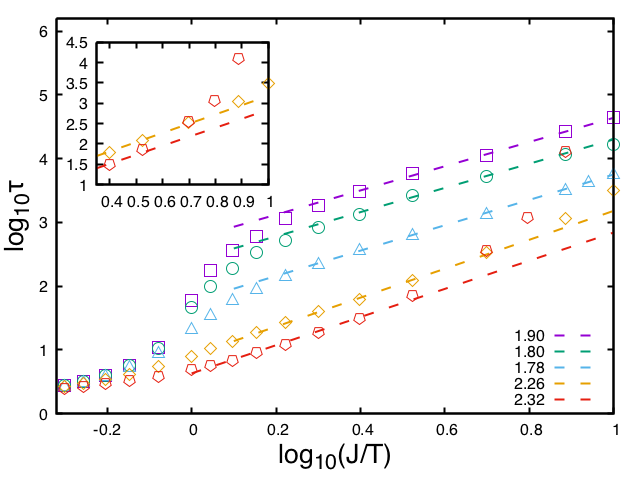}
\caption{Log-log plot of $\tau$ versus $J/T$. Same data as in Fig.~\ref{fig_arrhenius_tau}: $f=5/(34)^2, 10/(34)^2,1/34, 5/34, 13/34$ (from top to bottom). Dashed lines are linear fits to the data with slopes given in the figure. Inset: Zoom on the low-$T$ region for the two largest frustrations, $f=5/34$ and $f=13/34$.}
\label{fig_loglog_tau}
\end{figure}

This behavior qualitatively corresponds to what we have predicted in section \ref{sec_predictions}. The increase followed by a sub-Arrhenius dependence corresponds to the power-law temperature dependence predicted by Eq.~(\ref{eq_power_law}). To check this in more detail we display a log-log plot of  $\tau(T)$ in Fig.~\ref{fig_loglog_tau}. As shown in the figure, the intermediate temperature regime can indeed be fitted by a power-law behavior, but we find that the effective exponent extracted from the slope varies with frustration: it increases from about $1.9$ for the smallest frustration $f=5/(34)^2$, a value that is not too far from the exponent $3/2$ predicted in Eq.~(\ref{eq_power_law}), to $2.2-2.3$ for the largest ones. (The results are very similar when extracting $\tau$ from the other prescription: the values of the effective exponent are within $5\%$ of the stretched-exponential results.) 

We can rationalize the observed behavior as follows. For the smallest frustrations the activation energy is likely small enough that it is hard to disentangle Arrhenius from power-law dependence over the temperature range under study: recall that the estimate of the activation energy [see Eq.~(\ref{eq_arrhenius})] when $f\to 0^+$ is $0.19J$, so that the activated Arrhenius factor in $\tau(T)$ stays of O($1$) for $T\gtrsim0.1$. Actually, a fit to the last 3 points of the Arrhenius plot in Fig.~\ref{fig_arrhenius_tau}, for $5\leq J/T\leq 10$, gives an effective activation energy of the order of $0.24-0.28 J$ for the frustrations $f\leq 1/34$. For these frustrations  it then seems reasonable to assign the increase in the effective exponent of the power-law fit to the presence of a very smooth crossover to the Arrhenius regime.
 
Over the temperature range that we can access, {\it i.e.}, $T\geq0.1 J$, a crossover to an Arrhenius behavior is unambiguously detected only for the two largest frustrations, $f=5/34$ and $f=13/34$. This can be seen from the deviation from the power-law fit at low-$T$ in the inset of Fig.~\ref{fig_loglog_tau}, with a crossover temperature $T_{vl}$ around $0.2 J$ ($\log_{10}(J/T)\approx 0.7$) for $f=5/34$ and around $0.3J$ ($\log_{10}(J/T)\approx 0.5$) for $f=13/34$.  The range over which the power-law fit is good  appears to extend to lower temperature as $f$ decreases, so that the Arrhenius regime is too weak or  out of reach for $f\leq 1/34$.
 
\begin{figure}[h]
\includegraphics[width=\linewidth]{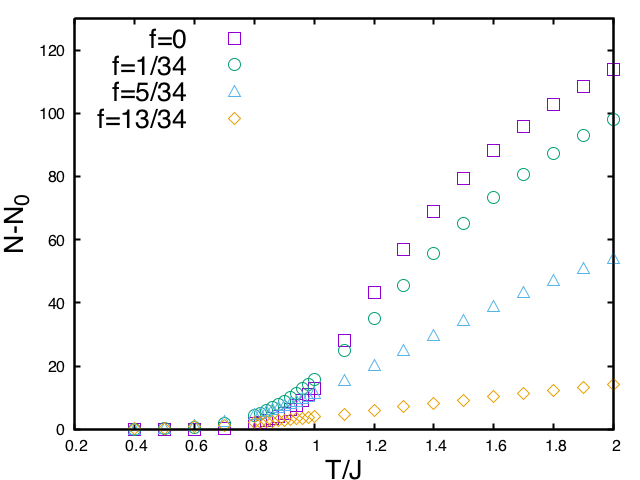}
\caption{Number of vortices $N$ minus the number of irreducible vortices $N_0$ for $f=1/34, 5/34, 13/34$,  and for $f=0$ with $L=40$ (the number of irreducible vertices is $34$,  $170$, $442$, and $0$, respectively). At $T\approx 0.7J$ and below the thermal vortices have virtually disappeared.}
\label{fig_vortices}
\end{figure}

Interestingly, the crossover $T_{vl}$ does not seem to correlate with the temperature at which virtually all thermal defects have disappeared and only the irreducible vortices induced by the flux $f$ remain. One can see from Fig.~\ref{fig_vortices} that this temperature is found rather around $T\approx 0.7J$, irrespective of the value of $f$. This temperature is better %more 
correlated with the establishment of the power-law regime (see Fig.~\ref{fig_loglog_tau} with $\log_{10}(0.7)\approx 0.155$).

For the two largest frustrations we find, as argued in section \ref{sec_predictions}, that the activation energy increases with $f$: it is equal to $0.45 J$ for $f=5/34$ and to $1.2 J$ for $f=13/34$ (and is larger than the estimate for small frustrations with $E\sim 0.2-0.3  J$).  This effect is similar to the increase of the activation energy found in a liquid with increasing density. Similar values of the order of $J$ have also been obtained in rather strongly frustrated XY models in the temperature regime below $T/J\approx 0.2$. The data of Kim and Lee\cite{kim-lee97} for $f=13/34$ can be fitted below $T\approx 0.25-0.2J$ with a low-temperature activation energy $E\approx 1.3 J$. (The authors try to describe the $T$-dependence by a super-Arrhenius dependence but as shown in Fig.~\ref{fig_arrhenius_tau} the super-Arrhenius character is not significant and tends to be mixed with the crossover with the power-law regime.) In the fully frustrated XY model ($f=1/2$), Granato\cite{granato12} observes Arrhenius behavior with $E\approx 1.0 J$, and in a related model of a curvature-induced frustrated XY model on a hyperbolic lattice with $f=-1/6$ (see also below), Baek {\it et al.}\cite{baek09} found $E\approx 0.92 J$.

Note finally that the curves appear to converge at high temperature above $T^*$: see Figs.~\ref{fig_arrhenius_tau} and \ref{fig_loglog_tau}. This is expected from the picture of a frustrated model being like an unfrustrated one in a finite size $L\sim \ell$. (For the two largest frustrations there are small deviations for temperatures higher than those shown in Figs.~\ref{fig_arrhenius_tau} and \ref{fig_loglog_tau},   %\IE{This deviation is not very visible in the figure. Should we include a separate figure? Or just say in words that there is a visible deviation upon zooming into the high T data?} 
but this merely reflects the fact that the characteristic length $\ell$, which is then equal to $1.6$ for $f=13/34$ and $2.6$ for $f=5/34$, is so small that  bulk behavior is not recovered.)

\subsection{Spin-wave kinetics versus activated vortex motion}
 
\begin{figure}[h]
\includegraphics[width=\linewidth]{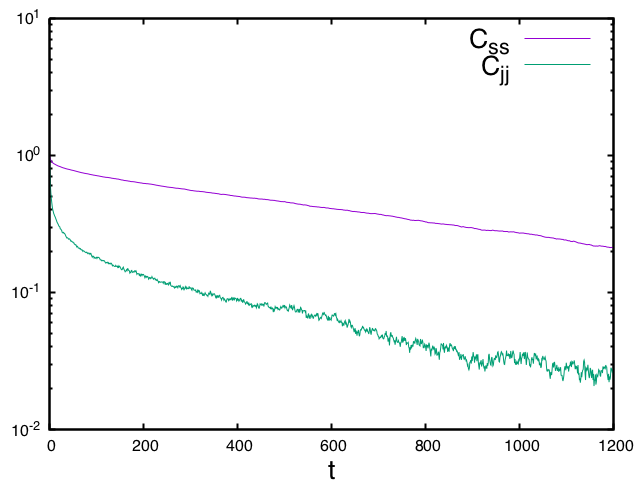}
\caption{Comparison of the time-dependent spin and current auto-correlation functions for $f=1/34$ and $T/J=0.3$. The latter has a faster initial decay but the relaxation times describing the final relaxation stage are comparable.}
\label{fig_cxy_jj}
\end{figure}

\begin{figure}[h]
\begin{subfigure}
  \centering
  \includegraphics[width=\linewidth]{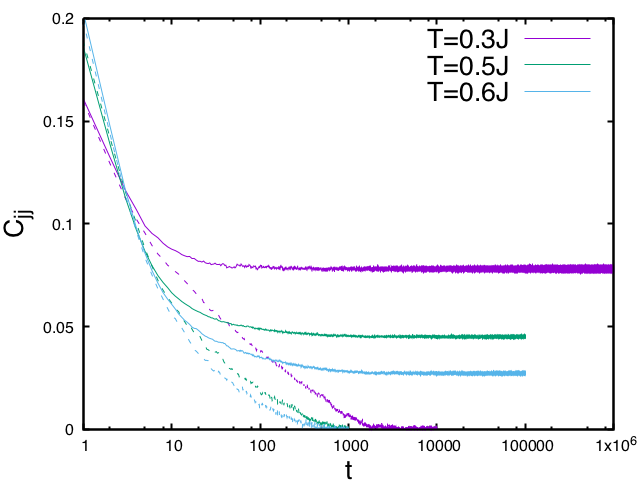}
  %\caption{(a)}
\end{subfigure}
\begin{subfigure}
  \centering
  \includegraphics[width=\linewidth]{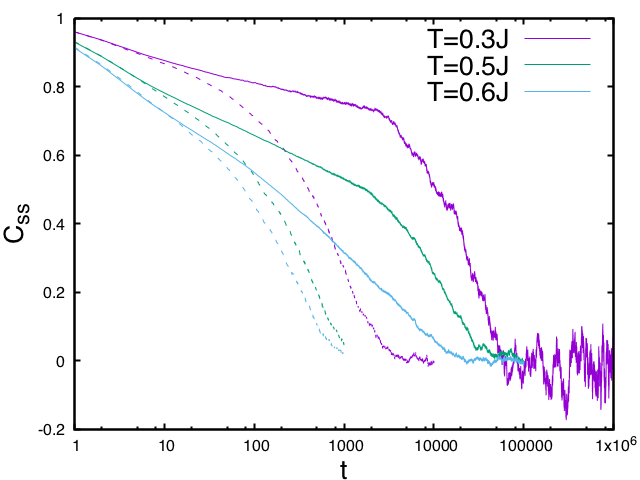}
  %\caption{(b)}
\end{subfigure}
\caption{Effect of pinning the frustration-induced vortices in the uniformly frustrated XY model for $f=1/34$ at several temperatures, $T/J=0.6, 0.5, 0.3$: Time-dependent current (top) and spin (bottom) auto-correlation functions in the system with and without vortex pinning. Solid lines are with vortex pinning and dashed lines are without.}
\label{fig_pinunpin}
\end{figure}

We have computed the current auto-correlation function $C_{jj}(t)$ to see whether the dynamics of the currents couples differently than that of the spins to the two relaxation mechanisms, respectively associated with spin waves and with the frustration-induced irreducible defects. One indeed expects that in the case of a complete decoupling between the spin waves and the defects (as in the Villain model\cite{villain75}), $C_{jj}(t)$ would mostly probe the activated motion of the irreducible vortices rather than the spin-wave kinetics and could be significantly slower than that of the spin autocorrelation function $C_{ss}(t)$. However, over the range of time and temperature that we could access, this is not what we have observed. As illustrated by the log-linear plot in Fig.~\ref{fig_cxy_jj} for $f=1/34$ at $T/J=0.3$, $C_{jj}(t)$  rather has a more rapid decay than $C_{ss}(t)$: this is apparent at short times (say for $t<100$ in Fig.~\ref{fig_cxy_jj}) while at longer times the two functions appear to decay more in parallel with however a slightly faster rate for $C_{jj}(t)$ (see also Fig.~\ref{fig_pinunpin} (a) and (b)). The absence of any additional slower relaxation in the current autocorrelation function is in line with what was found in Ref.~[\onlinecite{kim-lee97}], where the time-dependent vorticity auto-correlation function (directly sensitive to the vortices) and the spin auto-correlation function show a similar behavior.

To gain more insight, we have considered a system in which the frustration-induced vortices are pinned. We do so for $f=1/34$ at several temperatures, $T/J=0.7,\, 0.5,\, 0.3$, for which virtually all thermally induced vortices have disappeared (see Fig.~\ref{fig_vortices}). We have then computed both the current and the spin time-dependent autocorrelation functions.

In Fig.~\ref{fig_pinunpin} (a) we compare $C_{jj}(t)$ in the presence and in the absence of vortex pinning potential. The first rapid decay is similar in the two cases but, quite notably, the current auto-correlation function does not decay to zero in the presence of pinned vortices whereas it does in the unpinned case. (The height of the plateau at long time increases roughly linearly as temperature decreases, as one would expect as a result of the linear increase of the spin stiffness.) This clearly indicates that the long-time decay of $C_{jj}(t)$ in the unpinned case is due to the motion of the frustration-induced vortices. At the same time, as shown in Fig.~\ref{fig_pinunpin} (b), the spin autocorrelation function $C_{ss}(t)$ decays to zero whether vortices are pinned or not. The initial part of the decay is similar in the two cases but the final, and main, relaxation is significantly slower when the frustration-induced defects are pinned. Pinning the vortices therefore slows down the dynamics but does not prevent full relaxation of the spins.

These observations seem to confirm that in this temperature range, which according to Fig.~\ref{fig_loglog_tau} corresponds to a power-law temperature dependence of the relaxation time, the dynamics of the currents is controlled by the motion of the frustration-induced vortices whereas that of the spins is dominated by spin-wave kinetics. This makes the above finding about the comparable decay rate at long time time of $C_{jj}(t)$ and $C_{ss}(t)$ in the absence of vortex pinning even more surprising. This appears to imply that the two mechanisms at play, spin-wave kinetics and irreducible-vortex motion, have similar time scales in the temperature regime under consideration.

\section{Discussion}

We have found that the frustration-induced avoidance of the BKT %Berezinskii-Kosterlitz-Thouless 
transition in a uniformly frustrated $2$-$d$ XY model generates slow relaxation and complex dynamics in the system at temperatures near and below the avoided transition. However, the characteristics of this relaxation slowdown are qualitatively different than that found in glass-forming liquids. In particular, the super-Arrhenius temperature dependence of the relaxation time that is commonly observed in supercooled liquids is not reproduced by the present model (see the Introduction). Taken at face value, this means that small frustration and avoided criticality are not sufficient to generate the phenomenology associated with the glass transition. More ingredients are necessary.

Of special interest then is to contrast the properties of the $2$-$d$ uniformly frustrated XY model with those of atomic liquids in $2$-$d$ curved space.
The scenario of  ``frustration-avoided criticality''\cite{tarjus05,avoided1} is also realized by a one-component simple atomic liquid in $2$-dimensional curved space. %Glassiness in this case only takes place in curved space. 
With decreasing temperature, the liquid in Euclidean space easily goes into an ordered or quasi-ordered phase with $6$-fold symmetry: depending on the nature of the interaction potential, the liquid crystallizes in a hexagonal phase through one weakly first-order transition or through a sequence of two continuous or weakly first-order transitions separated by a narrow bond-orientationally ordered hexatic phase.\cite{nelson_book,krauth} However, curving space, {\it i.e.},  embedding the liquid in a $2$-$d$ manifold of constant curvature, thwarts crystallization. The prevalent local order has a $6$-fold symmetry (hexatic or hexagonal order) but its spatial extension is frustrated by the nonzero curvature,\cite{nelson_book,sausset} which then plays the role of the flux $f$ in the frustrated XY model. The transition or sequence of ordering transitions at or around a temperature $T^*$ is avoided because curvature imposes an irreducible density of topological defects, ``disclinations'' and ``dislocations'', in the underlying hexatic/hexagonal medium: the system stays in the liquid phase even below $T^*$ and only (possibly) encounters a defect-ordering phase transition at a much lower temperature where the irreducible defects form a lattice. One-component atomic liquids can therefore be ``supercooled'' by applying curvature and can subsequently form a glass.

These model glass-forming liquids in curved space have been investigated by analytical and numerical means both for constant negative curvature (the hyperbolic plane $H^2$ that is of infinite extent and cannot be embedded in $3$-d Euclidean space) \cite{sausset,sausset-nelson}Êand constant  positive curvature (the more familiar sphere $S^2$ of finite extent) \cite{vest}. It was found that crystallization is indeed avoided and that the dynamics of the ``supercooled'' liquid slows down as one lowers the temperature below $T^*$ at constant curvature. However, contrary to what we observe here for the uniformly frustrated XY model, the relaxation time apparently displays a super-Arrhenius temperature dependence. The ``fragility'', which characterizes how quickly the relaxation time and the transport coefficients increase with decreasing temperature,\cite{angell} changes with the curvature,  {\it i.e.}, the frustration: the more frustrated, the less fragile. Other characteristics of the relaxation slowdown in glass-forming liquids, such as the marked nonexponential time dependence of the relaxation functions and the increasingly heterogeneous character of the dynamics with the coexistence over an extended period of time of fast and slow regions,\cite{berthier-biroli11,tarjus-review} are also observed. 

Why are the dynamics of liquids in curved space and uniformly frustrated XY systems so different despite the fact that the spatial dimension of the manifold is the same? The frustration $f$ is of course not produced by the same mechanism in the two cases, being associated with the curvature in the former and with the flux in the latter. However, Baek {\it et al.}\cite{baek09} have  studied by Monte Carlo simulations the XY model on a hyperbolic lattice, in which a uniform frustration with $f=-1/6$ (corresponding to a typical frustration length $\ell\approx 2.45$) is generated by the curvature. The model shows no significant differences with the uniformly frustrated XY model in flat space with a similar frustration length ({\it i.e.}, our study for the largest frustrations). In particular, it does not display the super-Arrhenius behavior found for the liquid. 

The difference seems to rather stem from the nature of the degrees of freedom: spins in one case, with phase fluctuations ({\it i.e.}, spin waves) and vortex fluctuations; %, and 
particles in the other, with both translational and bond-orientational fluctuations and the associated defects, dislocations and disclinations. Disclinations are akin to vortices in the XY model. Dislocations, however, have been argued to play a crucial role in the physics of liquids in curved space at low enough temperature, including the relaxation via thermal activation.\cite{sausset-nelson} They have no direct counterparts in the uniformly frustrated XY model at low temperature. It is then possible that the latter system misses this important piece of liquid physics, the intertwining of translational and rotational degrees of freedom, and displays a relaxation slowdown associated with avoided criticality that is too dominated by spin-wave kinetics to be a minimal model for generic glass-forming liquids.

Further evidence pointing to a qualitative difference between the $2$-$d$ uniformly frustrated XY model and $2$-$d$ glass-forming liquids comes from the role of ``soft modes'' in the dynamics. The latter are the remains of the long wavelength excitations present in the absence of frustration, {\it i.e.}, spin waves in the XY model and density fluctuations in liquids, which have a drastic influence on the ordering behavior in $2$-dimensional systems.\cite{mermin-wagner} These soft modes have been recently shown to have a strong effect on the dynamics of $2$-$d$ glass-forming liquids.\cite{flenner15,shiba16,vivek17,illing17,tarjus17}  However, they can be disentangled from the more proper ``glassy'' component of the dynamics, which then appears similar to what is generically found in $3$-$d$ supercooled liquids.\cite{shiba16,vivek17,illing17,tarjus17} This is in contrast to what we have found here in the case of the uniformly frustrated XY model.

Based on the above discussion, one may then wonder whether there are generalizations of the $2$-$d$ uniformly frustrated models that are better suited for describing glass-forming liquids and yet still tractable, and, more generally, what are the additional ingredients to be added to   frustration-avoided criticality to produce a minimal theory of glass formation.

%These speculations raise two additional questions that would be worth addressing: Are there generalizations of the $2$-$d$ uniformly frustrated models that are better suited for describing glass-forming liquids and yet still tractable?  Is the phenomenon of frustration-avoided criticality relevant at all for glass formation?

Finally, it is worth mentioning that the uniformly frustrated $2$-$d$ XY model studied in this paper has experimental realizations, either in the form of arrays of Josephson junctions or of periodic networks of superconducting wires, and that the relaxation of the system can therefore be experimentally accessed as a function of both temperature and frustration. There have been many studies of this kind in the past,\cite{carini88,tighe91,yu92,vanderzant94,lachenmann94,harada96,baek04} focusing for instance on the resistive behavior of the system, but most of them focused on either the very low temperature regime where a transition to an ordered vortex phase may appear or to the immediate vicinity of the BKT transition. A consistent experimental description of the whole temperature range as a function of frustration, {\it i.e.}, magnetic flux, would therefore be of great interest.


\begin{thebibliography}{15}
\bibitem{footnote_gauge}
This is more subtle when considering a finite system with periodic boundary conditions, as we do here. There may then be an additional dependence on large-scale vorticity associated with loops wrapping around the system.\cite{alba08} For the local quantities that we consider in this work this dependence is however absent. In addition the periodic boundary conditions may frustrate the establishment of an ordered (or quasi-ordered) phase of irreducible vortices, but this is not relevant in the temperature domain under study where the system remains significantly above any transition to such ordered phases.
\bibitem{footnote_finitesize}
Note that we are not interested in establishing and studying the presence of a low-temperature (ordered, quasi-ordered, or glassy)  phase as we work at temperatures significantly above the putative transition: see Refs. [\onlinecite{franz-teitel95,hattel95}] for the small-frustration cases. For the stronger frustration $f=13/34$, which is an approximant to the irrational frustration given by the golden mean, the situation is more controversial.\cite{halsey85,granato96,kim-lee97,gupta98,park00} It has for instance been argued that there is a zero-temperature transition with associated finite-size effects.\cite{park00} However, for the temperature range and the system sizes that we study these effects are negligible.
\end{thebibliography}
\end{document}